# Tuning of Photoexcited Electron Dynamics at Monolayer h-BN/Metal Interfaces by Corrugation

*Masahiro Shibuta,*[a,b,*] *Maximilian Schaal,*[c] *Marco Gruenewald,*[c] *Jonas Brandhoff,*[c] *Felix Otto,*[c] *Roman Forker,*[c] *and Torsten Fritz*[c,*]

[a] Department of Physics and Electronics, Graduate School of Engineering, Osaka Metropolitan University, 1-1, Gakuen-cho, Naka-ku, Sakai, Osaka 599-8531, Japan

[b] Institute for Molecular Science, National Institutes of Natural Sciences, Okazaki, Aichi 444-8585, Japan

[c] Institute of Solid State Physics, Friedrich Schiller University Jena, Helmholtzweg 5, 07743 Jena, Germany

Address correspondance to

*Masahiro Shibuta; e-mail: shibuta@omu.ac.jp

*Torsten Fritz; e-mail: torsten.fritz@uni-jena.de

ABSTRACT

Atomic-scale corrugation in two-dimensional materials can modify interfacial electronic coupling, yet its influence on ultrafast carrier relaxation remains poorly established. Here, we compare image potential states (IPS) at monolayer h-BN/Ir(111) and h-BN/Pt(111) interfaces using structural characterization and time-resolved two-photon photoemission spectroscopy. Consistent with literature, h-BN is strongly corrugated on Ir(111) but comparatively flat on Pt(111). The first ($n$ = 1) and second ($n$ = 2) IPS appears at similar energies on both substrates, whereas their relaxation dynamics differ markedly. On h-BN/Ir(111), the IPS decay is response-limited ($<$20 fs), while h-BN/Pt(111) exhibits lifetimes of 56 fs ($n$ = 1) and 75 fs ($n$ = 2). The lifetime contrast is most consistently explained by corrugation-enhanced overlap of the IPS wavefunction with the metal substrate, which accelerates electron decay. These results indicate that atomic-scale corrugation is an effective physical parameter for tuning ultrafast electron dynamics at two-dimensional material and metal interfaces.

Two-dimensional (2D) layers supported on crystalline metal substrates provide atomically well-defined model interfaces in which the lattice registry, interfacial coupling, and vertical corrugation can be tuned through the overlayer/substrate combination.[1,2] In addition to their static electronic structures, such interfaces govern how photoexcited electrons exchange energy and charge with the underlying metal on femtosecond time scales. Understanding these ultrafast relaxation pathways is therefore essential for surface photochemistry, heterogeneous catalysis, carrier injection/extraction, and nanoscale energy dissipation processes.[3–6]

Monolayer hexagonal boron nitride (h-BN) forms well-ordered 2D moiré superstructures on transition-metal surfaces.[7,8] Owing to its wide band gap, h-BN serves as an atomically thin insulating spacer, while its periodic moiré lattice provides a template for the ordered arrangement of atoms, molecules, and nanoparticles.[7–9] The amplitude of the structural corrugation depends sensitively on lattice mismatch and on the strength of the electronic interaction with the metal substrate. In particular, previous diffraction and scanning probe studies have shown that h-BN exhibits pronounced vertical modulation on Ir(111), Rh(111), and Re(0001), whereas it remains comparatively flatter on Pt(111), Au(111), and Cu(111), reflecting a weaker interfacial interaction.[7,9–17] While the projected electronic band structures of Ir(111) and Pt(111) surfaces are broadly comparable near the Fermi level ($E_F$),[18–20] monolayer h-BN exhibits markedly different corrugation amplitudes on the two substrates because of the different lattice mismatches between h-BN and the respective metal surfaces (the height difference between strongly and weakly bound h-BN regions was reported as $\Delta h \approx 1.5$ Å on Ir(111)[15] versus $\Delta h \approx 0.5$ Å on Pt(111)[12]). This contrast allows us to examine the relationship between structural corrugation and interfacial electron dynamics.

Structural corrugation can modify the local surface potential, work function landscape, and effective tunneling barrier between an electron outside the surface and the metallic continuum.[1,2,7,8] It may therefore provide an efficient pathway for controlling the relaxation of excited electrons at 2D material/metal interfaces. However, the direct influence of substrate-induced corrugation on ultrafast excited-state relaxation has remained largely unexplored despite extensive studies of the geometric and static electronic structures of 2D material/metal systems.

Image potential states (IPS) are Rydberg-like unoccupied states bound outside a surface by the Coulombic image potential.[21–27] In an ideal IPS series, their energies are referenced to the vacuum level, $E_{vac}$, as $E_n = E_{vac} - (0.85 \text{ eV})/(n + a)^2$, where $n$ is the quantum number and $a$ is the quantum defect.[22–25] Because IPS wavefunctions extend from the surface into the vacuum while retaining finite overlap with the substrate, their lifetimes are extremely sensitive to changes in surface potential, tunneling barriers, and interfacial coupling. IPS therefore provide an excellent probe for examining how structural corrugation affects electron dynamics at surfaces and interfaces. IPS on monolayer h-BN/metal systems have been investigated in a number of scanning tunneling microscopy (STM) studies, particularly through field-emission resonance and *z*–*V* spectroscopy.[13,28–32] These measurements provide spatially resolved information on the energetic positions and wavefunction distributions of IPS. However, STM does not directly access the femtosecond relaxation dynamics of electrons that are transiently populated in IPS.

Two-photon photoemission (2PPE) spectroscopy using femtosecond laser pulses overcomes this limitation by directly populating and detecting unoccupied states in the time domain.[33–39] Although a few 2PPE studies have reported IPS formation and excited-electron dynamics at h-BN on metal surfaces (e.g., Ni(111)[40,41] and Cu(111)[42]), a quantitative and systematic assessment of how substrate-induced corrugation modifies the dynamics of excited electrons in IPS (IPS electrons) has not yet been established.

In this study, we combine structural and electronic characterizations with time-resolved 2PPE spectroscopy to compare the IPS formed at the monolayer h-BN on Ir(111) and Pt(111). The difference between both systems is the substrate-dependent corrugation: h-BN is strongly corrugated on Ir(111) ($\Delta h \approx 1.5$ Å),[15] whereas it is comparatively flat on Pt(111) ($\Delta h \approx 0.5$ Å).[12] We then show that, despite nearly identical IPS energies on the two substrates, the IPS relaxation dynamics are markedly different. These results demonstrate that atomic-scale corrugation is a decisive parameter governing interfacial electron dynamics at 2D material/metal interfaces.

We fabricated monolayer h-BN on clean Pt(111) and Ir(111) substrates through chemical vapor deposition,[43–45] and characterized the geometric structure using distortion-corrected[46] low-energy electron diffraction (LEED) (Figure 1). LEED patterns of h-BN/Ir(111) and h-BN/Pt(111) exhibit well-defined diffraction spots together with moiré-related multiple scattering spots, confirming the formation of ordered monolayer h-BN films on both substrates (Figures 1(a,b)). Quantitative analysis of the diffraction patterns yielded h-BN lattice constants close to 2.49 Å for both systems, consistent with monolayer h-BN.[7–9] The observed moiré periodicity of h-BN/Ir(111) is compatible with a 12-on-11 coincidence lattice, in which 12 h-BN unit cells match 11 Ir(111) surface unit cells. For h-BN/Pt(111), within the experimental uncertainty, the analyzed periodicity can be described by two closely related higher order commensurate approximants: 10 h-BN unit cells matching 9 Pt(111) surface unit cells, or 19 h-BN unit cells matching 17 Pt(111) surface unit cells. Details of the LEED analysis are provided in Note S1, Figure S1, Tables S1 and S2.

Although both systems form ordered moiré structures, the appearance of the moiré spots differs markedly between the two substrates (Figures 1(a,b) inset). For h-BN/Ir(111), the moiré spots are intense and sharply defined, indicating a pronounced periodic modulation of the surface structure and potential. In comparison, the moiré spots observed for h-BN/Pt(111) show slightly vaguer contrast and an azimuthal elongation, which is a result of slightly azimuthally rotated domains. The azimuthal disorder caused multiple minima in the potential landscape adopting in the growth process. The LEED result therefore indicates that the h-BN layer is less strongly locked to a unique registry on Pt(111). We emphasize that this azimuthal disorder is not explicitly considered in the epitaxial relations given above, since those refer exclusively to the center of the arcs. See the Supporting Information Note S1 for a more detailed discussion on this issue. These observations establish that the structural corrugation of monolayer h-BN depends sensitively on the underlying metal substrate, which is consistent with previous studies of h-BN/Ir(111) and h-BN/Pt(111);[9–12,15,16] the h-BN layer is more strongly corrugated on Ir(111) ($\Delta h \approx 1.5$ Å),[15] whereas it is comparatively flatter on Pt(111) ($\Delta h \approx 0.5$ Å).[12]

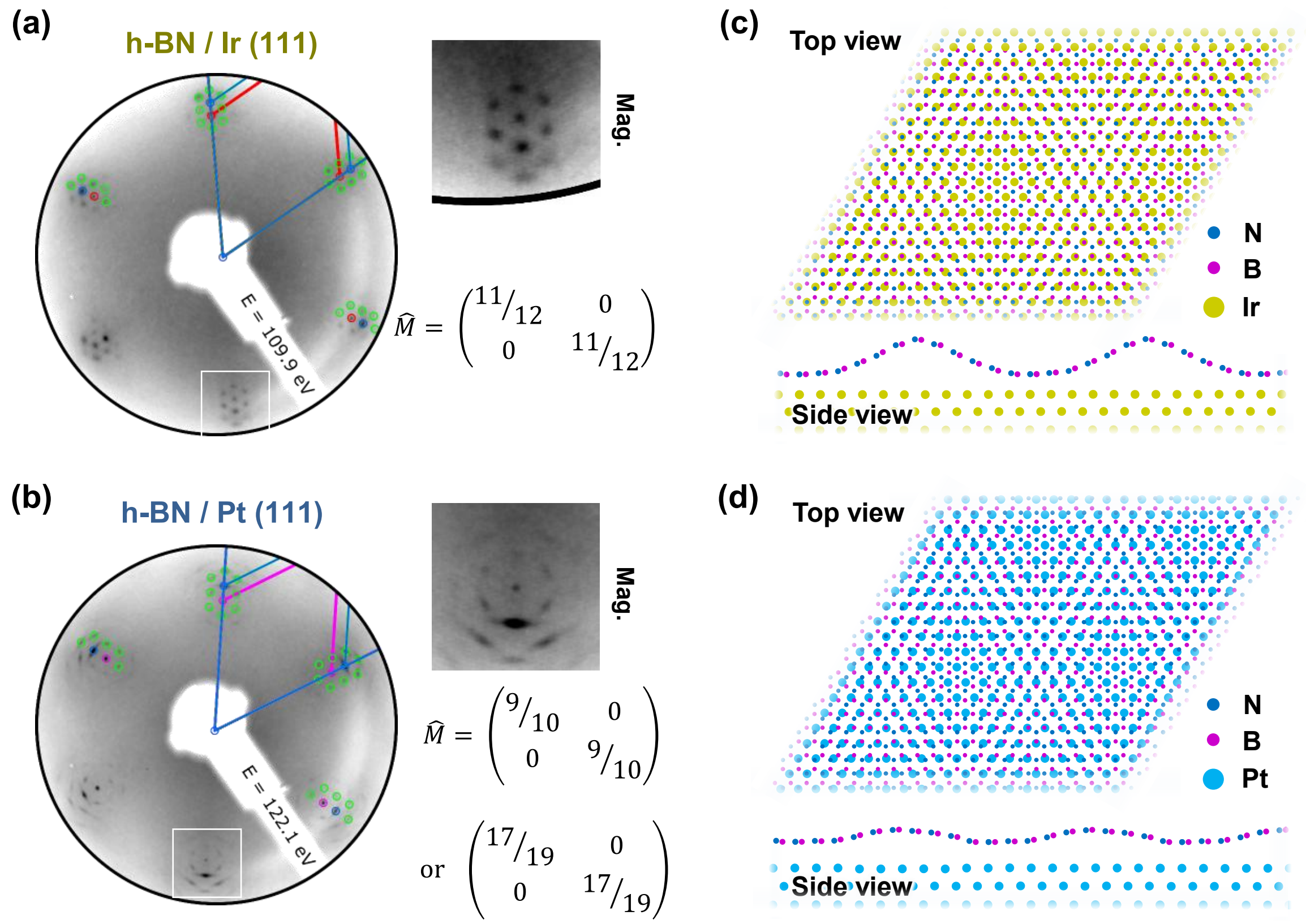

**Figure 1. Structural characterization and moiré registry of h-BN monolayers on Ir(111) and Pt(111).** (a,b) Distortion-corrected[46] LEED patterns of monolayer h-BN grown on (a) Ir(111) and (b) Pt(111), recorded at primary electron energies of $E_p$ = 109.9 eV and $E_p$ = 122.1 eV, respectively. The red, magenta, blue, and green markers denote substrate-derived diffraction spots, h-BN-derived spots, and multiple scattering spots, respectively. Quantitative LEED analysis yields the in-plane lattice constants of $|\vec{a}_1| = 2.488(3)$ Å and $|\vec{a}_2| = 2.490(3)$ Å for h-BN/Ir(111), and $|\vec{a}_1| = 2.487(3)$ Å and $|\vec{a}_2| = 2.490(3)$ Å for h-BN/Pt(111) (Table S2). The refined matrix notations are indicated in the figure, neglecting the present azimuthal disorder in the case of h-BN/Pt(111). Further details of the LEED analysis are provided in Note S1 and related supporting materials. Magnified diffraction spots are shown in the insets; the spots obtained from h-BN/Pt(111) are slight vaguer and have an azimuthal elongation. (c,d) Schematic top and side views of (c) h-BN/Ir(111) and (d) h-BN/Pt(111) systems. However, the depicted corrugations are not to scale.

Next, we characterized the local chemical environments and occupied electronic band structures of the h-BN films by X-ray photoelectron spectroscopy (XPS) and angle-resolved ultraviolet photoelectron spectroscopy (ARUPS). Figures 2(a,b) show the B 1s and N 1s core-level spectra of h-BN monolayer grown on Ir(111) and Pt(111). Both samples exhibited clear B 1s- and N 1s-derived peaks, and no XPS signal from unexpected elements (e.g., carbon and oxygen) is observed in the entire energy region (Figure S2), confirming the formation of pure h-BN layers on both substrates. The B 1s and N 1s spectra can be fitted by a dominant main component and a weaker shoulder on the higher binding energy side. The main B 1s and N 1s components of h-BN/Ir(111) are shifted toward higher binding energy compared with those of h-BN/Pt(111) (see Table S3 for fitted energy values), indicating a stronger electronic interaction between the h-BN layer and the Ir(111) substrate than the h-BN/Pt(111) case. [9–12,15,16]

The shoulder components observed at higher binding energies are attributed to the presence of B and N atoms in strongly interacting (valley) regions of the moiré unit cell, where the h-BN layer is closer to the metal substrate (Figures 1(c,d)). Such components have been reported for corrugated h-BN/metal systems and are commonly associated with registry-dependent local chemical environments. [9–12,15,16] The energy separation between the main and shoulder components is larger for h-BN/Ir(111) than for h-BN/Pt(111) (Table S3), suggesting a larger variation in the local chemical environment across the h-BN/Ir(111) moiré structure. These XPS results are consistent with previous observations: h-BN/Ir(111) exhibits a stronger registry-dependent interaction with the substrate, whereas h-BN/Pt(111) forms a comparatively weakly interacting and flatter monolayer. [9–12,15,16]

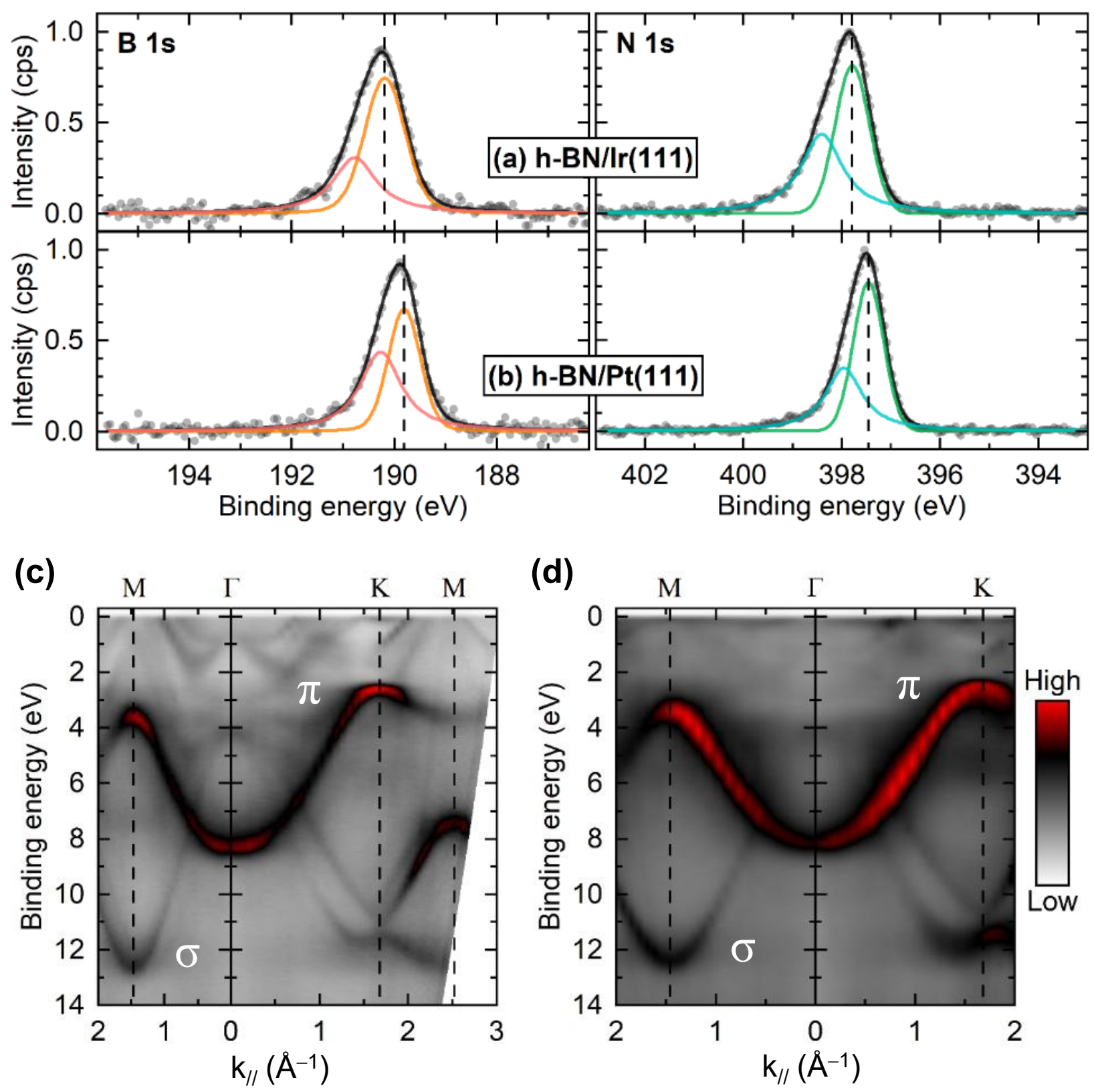

**Figure 2. Electronic characterization of the h-BN monolayers on Pt(111) and Ir(111).** (a,b) XPS data of monolayer h-BN on Ir(111) and Pt(111). Both B 1s (left) and N 1s (right) core level spectra are shown for (a) h-BN/Ir(111) and (b) h-BN/Pt(111). The black curves show the overall fits, while colored curves indicate the deconvoluted spectral components (Table S3). (c,d) ARUPS intensity maps ($h\nu$ = 40.8 eV) of the h-BN monolayer on (c) Ir(111) and (d) Pt(111). The intense dispersive signal around 3 – 8 eV in binding energy is attributable to the π-band of the h-BN layer, showing that both systems exhibit a similar π-band.

ARUPS measurements further characterize the occupied valence-band structure of the h-BN monolayers on both substrates. Figures 2(c,d) show the ARUPS intensity maps measured for (c) h-BN/Ir(111) and (d) h-BN/Pt(111), where a well-defined dispersive band is observed around 3 – 8 eV in both systems and is assigned to the well-known π valence band of monolayer h-BN.[7,8,47,48] The π band reaches its highest occupied energy near the K point and disperses toward higher binding energy around the Γ point, which is characteristic of single-layer h-BN. Note that the monolayer assignment is further supported by the characteristic π-band dispersion of h-BN, which differs from that of multilayer films, particularly near the K point.[49,50]

Importantly, the overall $\pi$-band dispersion is very similar for h-BN/Ir(111) and h-BN/Pt(111). This indicates that the occupied valence-band structure of the h-BN monolayer is largely preserved on both substrates, despite the different local adsorption geometries and corrugation amplitudes. Together with the XPS results, the UPS data show that the prepared h-BN films on Pt(111) and Ir(111) have comparable static occupied electronic structures, providing a suitable basis for comparing substrate-dependent IPS relaxation dynamics.

Having established the differences in corrugation-induced structural and electronic interactions, we examined the unoccupied electronic structure using 2PPE spectroscopy. Figure 3(a) shows the 2PPE spectra of h-BN monolayer on Ir(111) and Pt(111) substrates. The 2PPE spectra were obtained by a third harmonic of a titanium sapphire femtosecond laser, where the ultraviolet pulse (photon energy: $h\nu_{UV}$= 4.33 eV) was generated from the near-infrared (NIR) fundamental output. The bottom axis represents the kinetic energy ($E_{\mathrm{kin}}$) of the photoelectrons referenced to $E_{\mathrm{F}}$, indicating that the highest $E_{\mathrm{kin}}$ from the occupied states (i.e., Fermi edge) appears at $E_{\mathrm{F}}$ + 2 $h\nu_{UV}$. The top axis indicates the excited state energy with one-photon excitation with $h\nu_{UV}$; the Fermi edge should appear at $E_{\mathrm{F}}$ + 1 $h\nu_{UV}$. The workfunction, $\Phi$, ($\Phi$ = $E_{\mathrm{vac}} - E_{\mathrm{F}}$, determined from the leading edge of the secondary electron cutoff in $E_{\mathrm{kin}}$, are similar to each other: $\Phi$ = (4.70 ± 0.01) eV for h-BN/Pt(111) and $\Phi$ = (4.55 ± 0.01) eV for h-BN/Ir(111).

In both systems, a pronounced feature appears at approximately 4 eV above $E_{\mathrm{F}}$ in the excited state energy (top axis) and can be assigned to the first ($n$ = 1) IPS, which will be confirmed from multiple viewpoints (see below). The IPS peak on h-BN/Ir(111) at $E_{\mathrm{F}}$ + 3.84 eV appears somewhat broader than that on h-BN/Pt(111) at $E_{\mathrm{F}}$ + 3.92 eV. This broadening may reflect local variations of the workfunction[13,29] associated with the larger corrugation of h-BN/Ir(111).

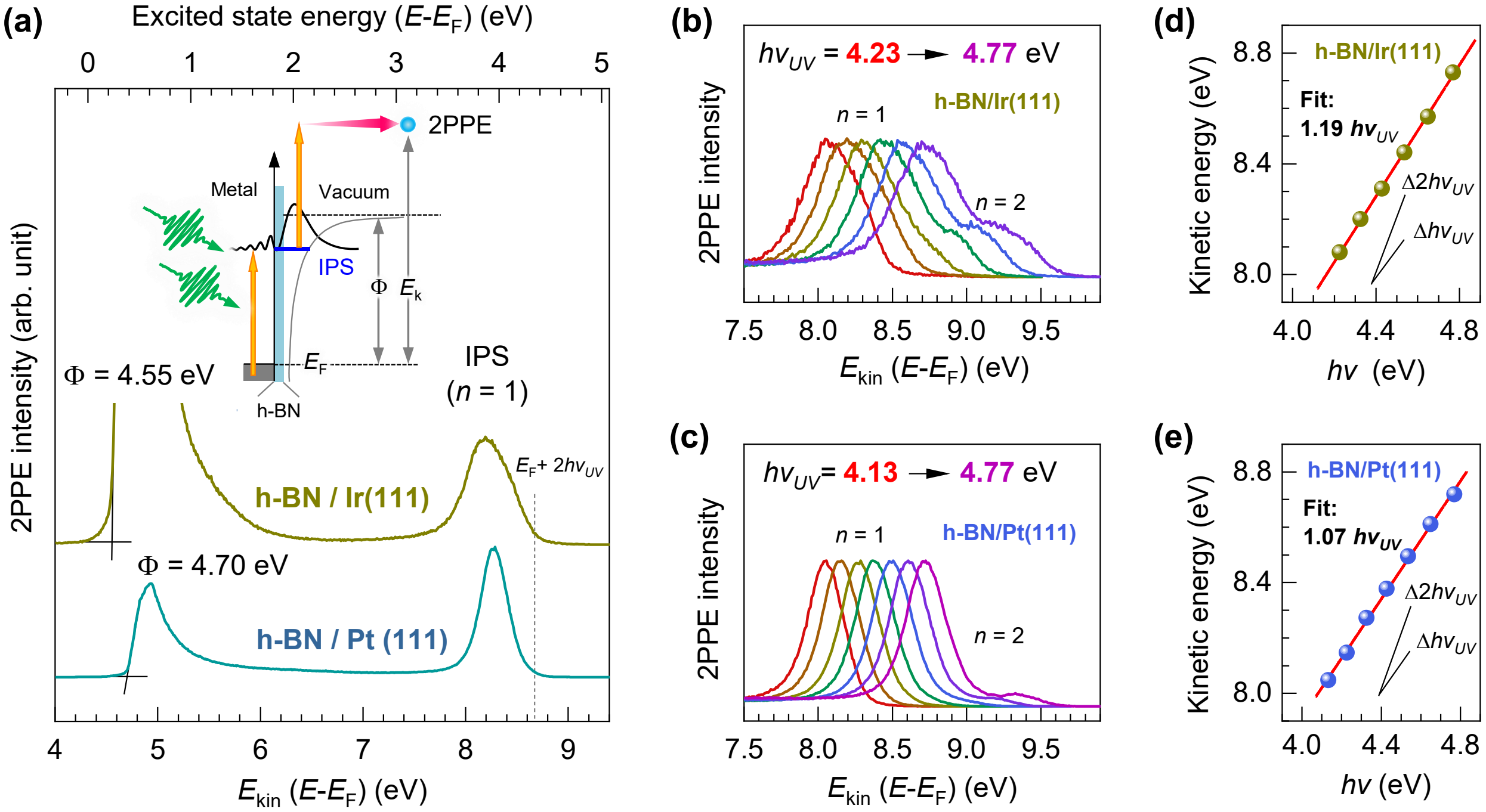

**Figure 3. Identification of IPS on h-BN monolayers grown on Ir(111) and Pt(111).** (a) 2PPE spectra ($h\nu_{UV}$ = 4.33 eV) for h-BN/Ir(111) and h-BN/Pt(111), showing the $n$ = 1 IPS on both systems. (b,c) 2PPE spectra recorded with various $h\nu_{UV}$ ranging 4.1 – 4.8 eV for (b) h-BN/Ir(111) and (c) h-BN/Pt(111), demonstrating systematic peak shifts depending on $h\nu_{UV}$. The higher-lying $n$ = 2 IPS appears at a higher $h\nu_{UV}$ range. (d,e) Extracted peak energy of $n$ = 1 IPS as a function of $h\nu_{UV}$. The linear dependence with a slope close to $\Delta h\nu_{UV}$ confirms the unoccupied-state character of the IPS.

To confirm that the observed peaks originate from an unoccupied intermediate state, 2PPE spectra were recorded while varying $h\nu_{UV}$ between 4.1 and 4.8 eV (Figures 3(b,c)). In both systems, the IPS peak shifts toward higher $E_{\mathrm{kin}}$ with increasing $h\nu_{UV}$. When the peak energy of $n$ = 1 IPS is plotted as a function of $h\nu_{UV}$ (Figures 3(d,e)), the shift closely follows the change in photon energy, $\Delta h\nu_{UV}$. This linear dependence indicates that the detected feature originates from an unoccupied intermediate state, consistent with the tentative excitation of electrons into the IPS by the first photon and subsequent photoemission by the second photon (Figure 3(a) inset). For h-BN/Ir(111), the extracted slope is slightly larger (1.19 $h\nu_{UV}$), suggesting that the observed signal may be influenced by additional contributions from occupied initial states.

Furthermore, using higher $h\nu_{UV}$, a weaker spectral feature in the 2PPE spectra was obtained at higher energy for both systems (Figures 3(b,c)), indicating that another higher-lying unoccupied state becomes accessible. From the energy position locating just below $E_{\mathrm{vac}}$ (i.e., at $E_{\mathrm{F}}$ + 4.38 eV for h-BN/Ir(111) and at $E_{\mathrm{F}}$ + 4.56 eV for h-BN/Pt(111), obtained by $E_{\mathrm{kin}} - h\nu_{UV}$), the peak is attributable to the IPS with the next quantum state ($n$ = 2). The comparatively stronger $n$ = 2 intensity on h-BN/Ir(111) may also be related to the interfacial interaction between h-BN and Ir(111), where the IPS wavefunction penetrates more into the metal substrate, resulting in an enhanced intermediate-state population or increased coupling with occupied states.

We further confirmed the IPS characteristics of the observed peaks using angle-resolved 2PPE measurements (Figure S3) and the light polarization dependence (Figure S4). The angle-resolved 2PPE reveals a free-electron-like parabolic dispersion for the $n$ = 1 IPS peak with an effective mass of (1.03 ± 0.06) × $m_e$ ($m_e$: rest mass of free electron) for h-BN/Ir(111), and comparable behaviors for both $n$ = 1 and $n$ = 2 IPS peaks were observed also for h-BN/Pt(111). The light polarization dependence for these spectral features exhibits complete selectivity for the p-polarized photon that includes an electromagnetic field component oriented normal to the surface. These spectral behaviors are indeed characteristic for IPS.[22–25] Therefore, the observed unoccupied state peaks indeed originate from $n$ = 1 and $n$ = 2 IPS formed on the h-BN monolayer grown on Ir(111) and Pt(111) exhibiting different corrugations.

Despite the clear differences in structural configuration and interfacial electronic interaction identified by LEED and XPS, the energetic positions and dispersion behaviors of both the π-band and IPSs are nearly identical in the h-BN/Pt(111) and h-BN/Ir(111) systems. These results demonstrate that the static occupied and unoccupied electronic structures, represented by the h-BN $\pi$-band and image-potential states, respectively, are largely preserved irrespective of substrate-induced corrugation.

In contrast to the nearly identical static electronic structures, time-resolved 2PPE measurements reveal a pronounced substrate dependence of the ultrafast relaxation dynamics of IPS electrons. The time-resolved measurements using different pump and probe photons are schematically shown in the inset of Figure 4(a). To obtain a better signal-to-noise ratio of the IPS signal, the IPS electron is pumped by $h\nu_{UV}$, and it is then probed by aftercoming NIR photons with $h\nu_{NIR}$. This pump-probe optical configuration results in the IPS peak appearing at different $E_{\mathrm{kin}}$ in the dual-color time-resolved 2PPE measurement (Figure 4(a)). Figures 4(b,c) show the time-resolved 2PPE data for the (b) h-BN/Ir(111) and (c) h-BN/Pt(111) systems obtained at various pump-probe delays. The excited state energies of IPSs (top axis in (b,c), calculated by $E_{\mathrm{kin}} - h\nu_{NIR}$) are identical to the single-color measurements (Figure 3).

Ultrafast relaxation of the IPS electrons is observed within a few hundred femtoseconds in these time-resolved 2PPE spectra. However, a closer comparison of the two datasets reveals different decay dynamics for the two substrates. Specifically, at a delay time of approximately 200 fs, the intensity of the $n$ = 1 IPS on h-BN/Ir(111) decreases to approximately one-third of its value at a delay time of ~100 fs. In contrast, on h-BN/Pt(111), the intensity decreases to only a half over the same delay interval. Furthermore, a finite IPS signal is still observable at a delay time of approximately 300 fs for the h-BN/Pt(111), while it already has completely disappeared for h-BN/Ir(111). These observations indicate that the IPS electrons on h-BN/Pt(111) possess a longer lifetime than those on h-BN/Ir(111). The remarkable difference in both systems, despite nearly identical energetic and dispersion characteristics, demonstrates the corrugation-dependent control of IPS electron dynamics.

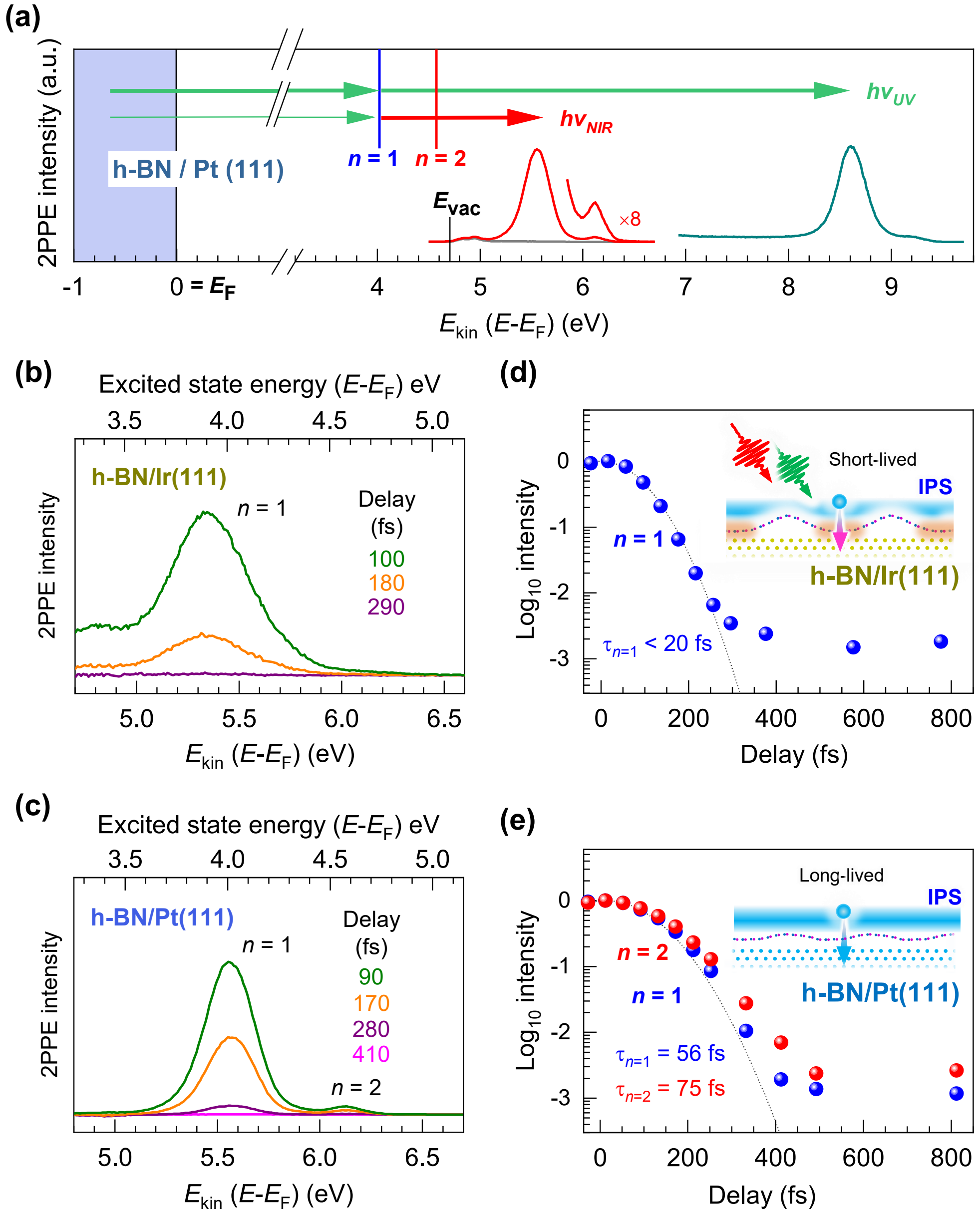

**Figure 4. Time-resolved relaxation dynamics of image potential states.** (a) Excitation schemes using $2 \times h\nu_{UV}$ for resolving IPS and using $h\nu_{UV}$ and $h\nu_{NIR}$ for time-resolved 2PPE. $E_{\text{kin}}$ of the photoelectron from the IPS shift by the $h\nu$ difference between both optical configurations ($h\nu_{UV}$ + $h\nu_{UV}$) − ($h\nu_{UV}$ + $h\nu_{NIR}$). (b,c) Time-resolved 2PPE spectra for (b) h-BN/Ir(111) and (c) h-BN/Pt(111) recorded with delay times between $h\nu_{UV}$ and $h\nu_{NIR}$ after arriving $h\nu_{UV}$. The intensity decrease at similar delay times is clearly more pronounced for h-BN/Ir(111) than for h-BN/Pt(111), indicating a longer lifetime of the IPS electron for the latter system. (d,e) Intensity traces of the IPS signals in (b) and (c). The decay occurs within the system response function, RF (185 fs in FWHM) for (d) h-BN/Ir(111), while the decay curve deviates from the RF for (e) h-BN/Pt(111).

Convolution fitting yields IPS electron lifetimes of (56 ± 23) fs ($n$ = 1) and (75 ± 19) fs ($n$ = 2) for h-BN/Pt(111), whereas they were much shorter on h-BN/Ir(111), being at most 20 fs. The insets in (d) and (e) illustrate the corrugation-induced control of the IPS electron dynamics.

To quantitatively evaluate the difference in IPS dynamics, the IPS intensities were plotted as a function of pump–probe delay time, as shown in Figures 4(d) and (e). For h-BN/Ir(111), the intensity trace of the IPS mostly overlaps with the system response function (RF) curve (width: 185 fs in FWHM), indicating a significantly fast relaxation of the IPS electron. From the absence of resolvable deviation from the RF curve, the lifetime of the IPS electron is estimated to be shorter than 20 fs. In contrast, the intensity traces of IPS ($n$ = 1 and 2) obviously deviate for h-BN/Pt(111) from the RF curve. The intensity traces were then analyzed using a convolution of the RF and a single-exponential decay model, yielding the lifetimes of IPS electrons to be (56 ± 23) fs for $n$ = 1 and (75 ± 19) fs for $n$ = 2. The longer lifetime in the $n$ = 2 IPS is a common behavior in IPS studies, as the IPS wavefunction for $n$ = 2 extends farther away from the surface than for $n$ = 1.

Regarding the IPS electron dynamics for clean metal substrates, for Pt(111), it has been reported that the lifetimes of IPS electrons are (26 ± 7) fs for $n$ = 1 and (62 ± 7) fs for $n$ = 2.[51] Compared with the change in the IPS lifetime by the formation of an insulating monolayer with rare-gas atoms (Ar, Xe, Kr) on noble metal substrates,[23–25,35,52] the increment of IPS lifetimes at the h-BN monolayer is comparable including the quantitative viewpoint; the van der Waals thickness of h-BN (~3.3 Å)[1,2] is comparable to the thickness of rare-gas monolayers (3.6 - 4.3 Å) showing IPS lifetimes of <100 fs on most systems.[23–25]

On the other hand, no direct lifetime measurement has been reported so far for the IPS on clean Ir(111) to the best of our knowledge, whereas it has been spectroscopically resolved recently by multi-photon photoemission (5PPE).[53] Because IPS decay can involve several channels, including electron–hole pair excitation, electron-phonon and electron-plasmon scatterings, and other many-body processes, its lifetime cannot be inferred solely from the one-electron band structure.[22–25,35] Nevertheless, the broadly comparable surface-projected electronic structures of Ir(111) and Pt(111) [18–20,51,53] provide no clear indication that the clean Ir(111) IPS should exhibit an intrinsically exceptional lifetime. If the h-BN overlayer simply acted as a flat insulating spacer with comparable decoupling efficiency on both Ir(111) and Pt(111), a measurable IPS lifetime longer than 20 fs would also be expected on h-BN/Ir(111). The response-limited fast decay observed in h-BN/Ir(111) therefore suggests that the highly corrugated h-BN/Ir(111) layer introduces an additional relaxation pathway by enhancing the coupling of the IPS wavefunction to the Ir(111) substrate.

For clean and atom/molecule-covered metal surfaces, IPS lifetimes are known to decrease strongly when the IPS wavefunction penetrates more deeply into the substrate and overlaps more efficiently with bulk electronic states.[23–25,34,35] In a strongly corrugated h-BN layer (e.g., h-BN/Ir(111)), local variations in the vertical position of the overlayer and in the electrostatic potential can enhance the IPS wavefunction overlap with the substrate and reduce the effective tunneling barrier in specific regions of the moiré unit cell. This interpretation is consistent with STM/STS studies of corrugated h-BN/metal moiré structures, where field-emission-resonance features and the associated local workfunction contrast vary strongly within the moiré unit cell rather than simply following the geometric corrugation characterized by X-ray standing wave measurements.[15,29] The significantly shorter lifetime on h-BN/Ir(111) is therefore most consistently attributed to the corrugation-enhanced coupling of the IPS to the Ir(111) substrate.

The comparative 2PPE measurements demonstrate that atomic-scale structural corrugation directly influences ultrafast excited-state dynamics at 2D material/metal interfaces. Even when the static electronic structure appears nearly identical, subtle differences in the physical interfacial structure can strongly modify the ultrafast dynamics of IPS electrons.

In summary, we have investigated ultrafast dynamics of IPS electrons at a monolayer of h-BN on Ir(111) and Pt(111) by time-resolved 2PPE spectroscopy, where the structural and electronic characterizations were carried out using LEED/XPS/UPS. While the static electronic structures, including IPS energies, are nearly identical between h-BN/Ir(111) and h-BN/Pt(111), the relaxation dynamics of the IPS electrons differ markedly; the IPS electrons on h-BN/Ir(111) decay much faster than those on h-BN/Pt(111). The substantially shorter lifetime of h-BN/Ir(111) correlates with its larger structural corrugation and stronger interfacial interaction. These findings demonstrate that atomic-scale corrugation, even in atomically thin insulating overlayers, can directly govern ultrafast excited-state relaxation. Our results establish structural modulation as an effective parameter for controlling the excited electron dynamics at 2D material/metal interfaces.

## ASSOCIATED CONTENT

### Supporting Information

The experimental details, additional LEED analysis of the moiré periodicities (Note S1, Figure S1, and Tables S1, S2), XPS peak energies (Table S3), XPS survey spectra (Figure S2), angle-resolved 2PPE (Figure S3), and polarization-dependent 2PPE data (Figure S4).

## AUTHOR INFORMATION

### Corresponding Authors

*Masahiro Shibuta

e-mail: shibuta@omu.ac.jp

*Torsten Fritz

e-mail: torsten.fritz@uni-jena.de

**Author Information**

Masahiro Shibuta − Department of Physics and Electronics, Graduate School of Engineering, Osaka Metropolitan University, 1-1, Gakuen-cho, Naka-ku, Sakai, Osaka 599-8531, Japan, and Institute for Molecular Science, National Institutes of Natural Sciences, Okazaki, Aichi 444-8585, Japan

Maximilian Schaal − Institute of Solid State Physics, Friedrich Schiller University Jena, Helmholtzweg 5, 07743 Jena, Germany

Marco Gruenewald − Institute of Solid State Physics, Friedrich Schiller University Jena, Helmholtzweg 5, 07743 Jena, Germany

Jonas Brandhoff − Institute of Solid State Physics, Friedrich Schiller University Jena, Helmholtzweg 5, Jena 07743, Germany

Felix Otto − Institute of Solid State Physics, Friedrich Schiller University Jena, Helmholtzweg 5, Jena 07743, Germany

Roman Forker − Institute of Solid State Physics, Friedrich Schiller University Jena, Helmholtzweg 5, Jena 07743, Germany; orcid.org/0000-0003-0969-9180; Email: roman.forker@uni-jena.de

Torsten Fritz − Institute of Solid State Physics, Friedrich Schiller University Jena, Helmholtzweg 5, Jena 07743, Germany

**Author Contributions**

M. Shibuta and T.F. conceived the project. M. Shibuta, M. Schaal, M.G., J.B., and F.O. prepared the h-BN/metal samples and performed the LEED, XPS, UPS, and 2PPE measurements and data analysis. R.F. and T.F. contributed to the interpretation of the experimental data. M. Shibuta wrote the first version of the manuscript, which was finalized through discussions with all authors. All authors have approved the final version of the manuscript.

**Notes**

The authors declare no competing financial interest.

**ACKNOWLEDGMENT**

This work was supported by the Mitsubishi Foundation, the Murata Science Foundation, the Asahi Glass Foundation, the MEXT Leading Initiative for Excellent Young Researchers (No. JPMXS0320220123), JSPS Grant-in-Aid for Challenging Research (Pioneering) (No. 22K18268), JSPS Grants-in-Aid for Scientific Research (Nos. 26K08190, 24K01277, 23H01939), JSPS Fostering Joint International Research (No. 24KK0257) and JSPS Grant-in-Aid for Transformative Research Areas (A) (No. 26A203). We also gratefully acknowledge the financial support from the Deutsche Forschungsgemeinschaft through grant no. FO 770/3-1. J.B. acknowledges financial support through a PhD scholarship from the Landesgraduiertenstipendium funded by the State of Thuringia. The authors used ChatGPT by OpenAI to improve English language clarity and readability during manuscript preparation and to assist in the conceptualization and preparation of the TOC graphic. The authors reviewed and edited all AI-assisted outputs and take full responsibility for the final content.

*Supporting Information for*

# Tuning of Photoexcited Electron Dynamics at Monolayer h-BN/Metal Interfaces by Corrugation

*Masahiro Shibuta,*[a,b,*] *Maximilian Schaal,*[c] *Marco Gruenewald,*[c] *Jonas Brandhoff,*[c] *Felix Otto,*[c] *Roman Forker,*[c] *and Torsten Fritz*[c,*]

[a] Department of Physics and Electronics, Graduate School of Engineering, Osaka Metropolitan University, 1-1, Gakuen-cho, Naka-ku, Sakai, Osaka 599-8531, Japan

[b] Institute for Molecular Science, National Institutes of Natural Sciences, Okazaki, Aichi 444-8585, Japan

[c] Institute of Solid State Physics, Friedrich Schiller University Jena, Helmholtzweg 5, 07743 Jena, Germany

*E-mail: shibuta@omu.ac.jp, torsten.fritz@uni-jena.de

**Table of contents**

## Experimental details

### General

All experiments were carried out under ultrahigh-vacuum (UHV) conditions with a base pressure in the $1\times10^{-10}$ mbar range, including the preparation chamber, low-energy electron diffraction (LEED), reflection high-energy electron diffraction (RHEED), X-ray photoelectron spectroscopy (XPS), ultraviolet photoelectron spectroscopy (UPS), and two-photon photoemission (2PPE) spectroscopy. While sample preparation with *in-situ* RHEED, and 2PPE spectroscopy are performed in one UHV system, LEED, XPS, and UPS are available in a different UHV system.

The sample transfer between the two UHV systems is performed using a portable UHV system pumped by a battery-driven ion getter pump, where the vacuum pressure was kept below $1\times10^{-9}$ mbar during the transfer process. The samples were therefore not exposed to air during preparation and characterization procedures. We also cross-checked that no structural or spectral change was observed before and after each measurement. All LEED, XPS, UPS, and 2PPE measurements were carried out at room temperature.

### Preparation of monolayer h-BN on Pt(111) and Ir(111)

Prior to h-BN growth, the Pt(111) and Ir(111) single crystals were prepared by repeated cycles of $Ar^+$ ion sputtering and subsequent annealing at 800 °C. The cleanliness and the long-range ordering of the (111) termination of the substrates were verified by LEED, XPS, and UPS measurements.

Monolayer h-BN was grown by chemical vapor deposition (CVD) using borazine, $B_3N_3H_6$, as the molecular precursor.[1,2] For h-BN/metal, borazine was dosed onto the hot Ir(111) and Pt(111) substrates at a substrate temperature of 850 °C, using a borazine pressure in the low $10^{-8}$ mbar range and an exposure sufficient to reach saturated monolayer coverage. The growth of well-ordered h-BN layers was confirmed by *in-situ* imaging of RHEED patterns. The formation and quality of the h-BN monolayers were also checked by LEED and photoelectron spectroscopy. The absence of detectable carbon and oxygen signals after preparation was confirmed by XPS (Figure S2).

### LEED

LEED measurements were performed using LEED optics equipped with double-microchannel plate detectors (OCI Vacuum Microengineering). The recorded diffraction images were corrected for geometrical distortions and electron energy deviations using the LEEDCal software.[3] After calibration, the diffraction patterns were analyzed with the LEEDLab software to determine the lattice parameters of the corresponding diffraction spots.

**XPS/UPS**

XPS and UPS measurements were carried out using monochromatized Al Kα ($h\nu$ = 1486.7 eV, SPECS Focus 500) and unmonochromatized He II ($h\nu$ = 40.8 eV, SPECS; UVS 10/35) radiations, respectively (SPECS; UVLS). Photoelectrons were analyzed using hemispherical electron analyzers (SPECS PHOIBOS 150), providing overall energy resolutions of approximately 0.5 eV for XPS and 60 meV for UPS.

**2PPE Measurements**

2PPE experiments were performed using a tunable titanium sapphire laser system (Coherent Mira, 740 - 920 nm, ~100 fs, 76 MHz). In the single-color configuration for the identification of IPS (Figure 3 in the main text), the third harmonic (4.04 – 5.03 eV) of the fundamental output served as the excitation source as $h\nu_{UV}$. The laser beam was focused onto the sample surface by an aluminum concave mirror ($f$ = 300 mm), with an incidence angle of 30° relative to the surface normal. Photoelectrons emitted along the surface normal were analyzed using a hemispherical electron energy analyzer with a mean electron pathway radius of 200 mm.

For time-resolved 2PPE measurements (Figure 4), a pump–probe scheme was implemented by introducing both NIR fundamental ($h\nu_{NIR}$) and $h\nu_{UV}$ through an optical delay stage in a dual-color configuration. The overall energy resolution of the 2PPE setup was approximately 100 meV, and the temporal resolution was about 20 fs. Unless otherwise noted, incident photons were p-polarized; polarization-dependent measurements were performed by rotating the polarization using a half-waveplate. We carefully checked that no sample degradation occurred during the entire 2PPE experiments, which was also confirmed by LEED, XPS, and UPS.

**Note S1. LEED analysis and epitaxial relations**

In the main text, we showed the LEED images for both h-BN/Ir(111) and h-BN/Pt(111) systems, and we briefly explained the formation of moiré superstructures because of lattice mismatches (Figure 1). Here we describe the detailed result of the quantitative analysis. Lattice parameters of the substrates used in the analysis are summarized in Table S1.[4] The magnified LEED images of both systems are shown in Figure S1 (corresponding to Figure 1 in the main text) in which the LEED images are contrast-inverted and depicted with a logarithmic intensity scale. Diffraction spots of the Ir(111)/Pt(111) substrates, the h-BN layer, and multiple scattering spots are marked in red/magenta, blue, and green, respectively. The extracted parameters from the LEED image and their refined parameters are summarized in Table S2.

Considering the experimentally determined lattice constants and their uncertainties, the moiré superstructures arising from the lattice mismatch between the h-BN layer and the Ir(111) or Pt(111) surfaces were described using refined commensurate epitaxy matrices. For h-BN/Ir(111), the LEED analysis supports a coincidence structure in which 12 h-BN unit cells match 11 Ir(111) surface unit cells. For h-BN/Pt(111), the experimental accuracy does not allow to distinguish unambiguously between two closely related higher order commensurate approximants, namely 10 h-BN unit cells matching 9 Pt(111) surface unit cells and 19 h-BN unit cells matching 17 Pt(111) surface unit cells. The analysis is further complicated by the presence of azimuthal disorder in this case, which expresses itself as arcs rather than sharp spots in the LEED images. For the sake of simplicity, we give the epitaxial relations without taking into account the azimuthal disorder, considering only the centers of the arcs where the intensity is highest. Therefore, the epitaxial matrices given represent only one of many epitaxies, differing in the azimuthal angle $\delta$ in Table S2. Note that the lengths of the lattice vectors are not affected by the azimuthal disorder.

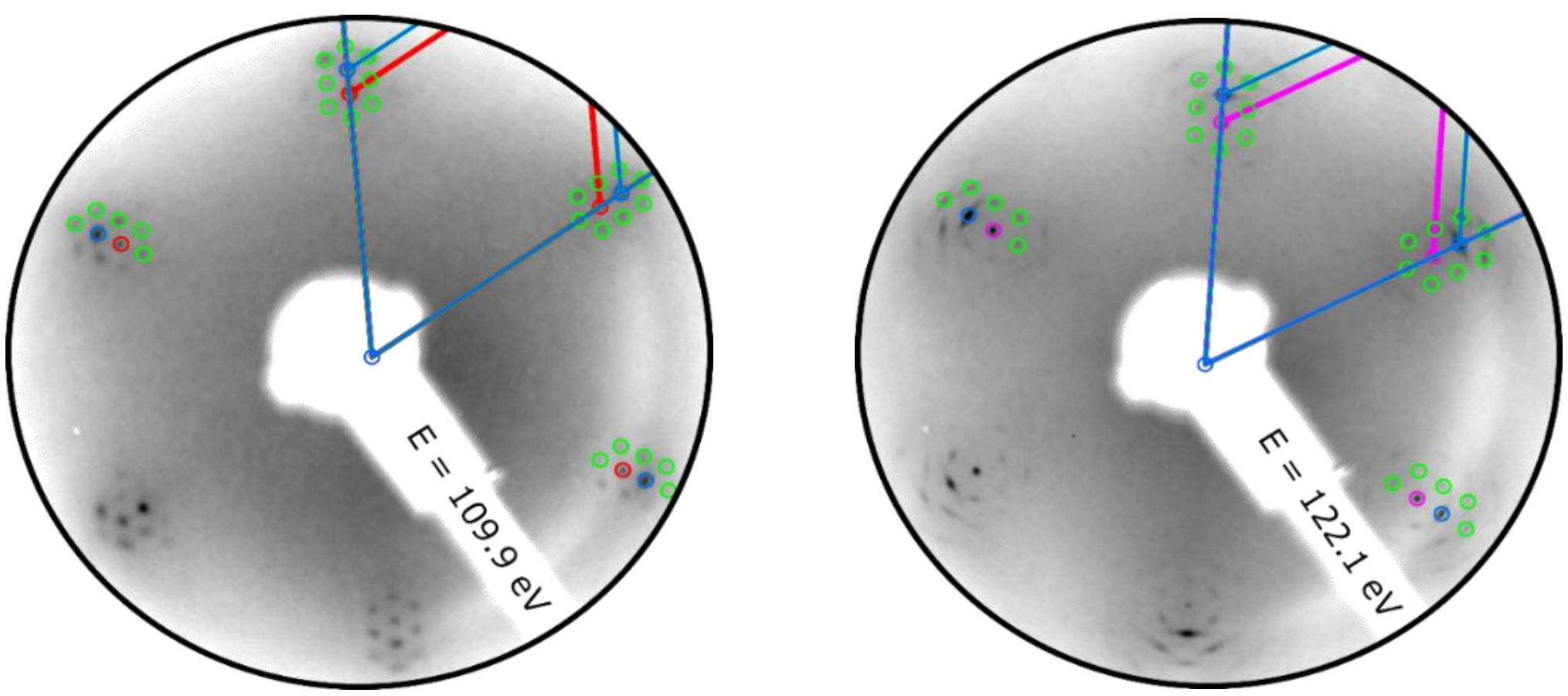

**Figure S1. Magnified LEED images of monolayer h-BN on Ir(111) and Pt(111).** LEED patterns of (left) h-BN/Ir(111) and (right) h-BN/Pt(111), recorded at primary electron energies of $E_p$ = 109.9 eV and $E_p$ = 122.1 eV, respectively.

**Table S1. Lattice parameters used in the LEED analysis.**[4]

| | Lattice constant of the substrate: $a$ (Å) | Surface lattice constant: $a_{[111]}$ (Å) |
|---|---|---|
| Ir(111) | 3.8392 | 2.7147 |
| Pt(111) | 3.9236 | 2.7744 |

**Table S2. Experimentally determined and refined lattice parameters.** The indicated uncertainties are the statistical errors of the lattice fit, given as $4\sigma$ intervals. The refined epitaxy matrices represent the closest higher-order commensurate approximants, where the rational fractions exhibit reasonably small denominators. From these approximants, the corresponding lattice parameters were calculated using the $a_{[111]}$ values provided in Table S1.

| | Epitaxy matrix: $\hat{M}$ (exp.) | Lattice parameters (exp.) | Refined epitaxy matrix: $\hat{M}$ | Refined lattice parameters: |
|---|---|---|---|---|
| h-BN /Ir(111) | $\begin{pmatrix} 0.917(4) & 0.000(4) \\ 0.001(4) & 0.918(4) \end{pmatrix}$ | $\lvert\vec{a}_1\rvert = 2.49(1)$ Å<br>$\lvert\vec{a}_2\rvert = 2.49(1)$ Å<br>$\gamma = 120.0(3)°$<br>$\delta = 0.0(2)°$ | $\begin{pmatrix} 11/12 & 0 \\ 0 & 11/12 \end{pmatrix}$ | $\lvert\vec{a}_1\rvert = 2.4885$ Å<br>$\lvert\vec{a}_2\rvert = 2.4885$ Å<br>$\gamma = 120°$<br>$\delta = 0°$ |
| h-BN /Pt(111) | $\begin{pmatrix} 0.896(4) & 0.000(3) \\ 0.001(4) & 0.898(4) \end{pmatrix}$ | $\lvert\vec{a}_1\rvert = 2.49(1)$ Å<br>$\lvert\vec{a}_2\rvert = 2.49(1)$ Å<br>$\gamma = 120.0(3)°$<br>$\delta = 0.0(2)°$ | $\begin{pmatrix} 9/10 & 0 \\ 0 & 9/10 \end{pmatrix}$ | $\lvert\vec{a}_1\rvert = 2.4970$ Å<br>$\lvert\vec{a}_2\rvert = 2.4970$ Å<br>$\gamma = 120°$<br>$\delta = 0°$ |
| | | | $\begin{pmatrix} 17/19 & 0 \\ 0 & 17/19 \end{pmatrix}$ | $\lvert\vec{a}_1\rvert = 2.4824$ Å<br>$\lvert\vec{a}_2\rvert = 2.4824$ Å<br>$\gamma = 120°$<br>$\delta = 0°$ |

**Table S3. Peak position parameters of XPS in binding energy.**

| | B 1s main (eV) | B 1s shoulder (eV) | N 1s main (eV) | N 1s shoulder (eV) |
|---|---|---|---|---|
| h-BN/Ir(111) | 190.26(6) | 190.85(14) | 397.86(2) | 398.48(5) |
| h-BN/Pt(111) | 189.92(8) | 190.38(17) | 397.57(2) | 398.07(6) |

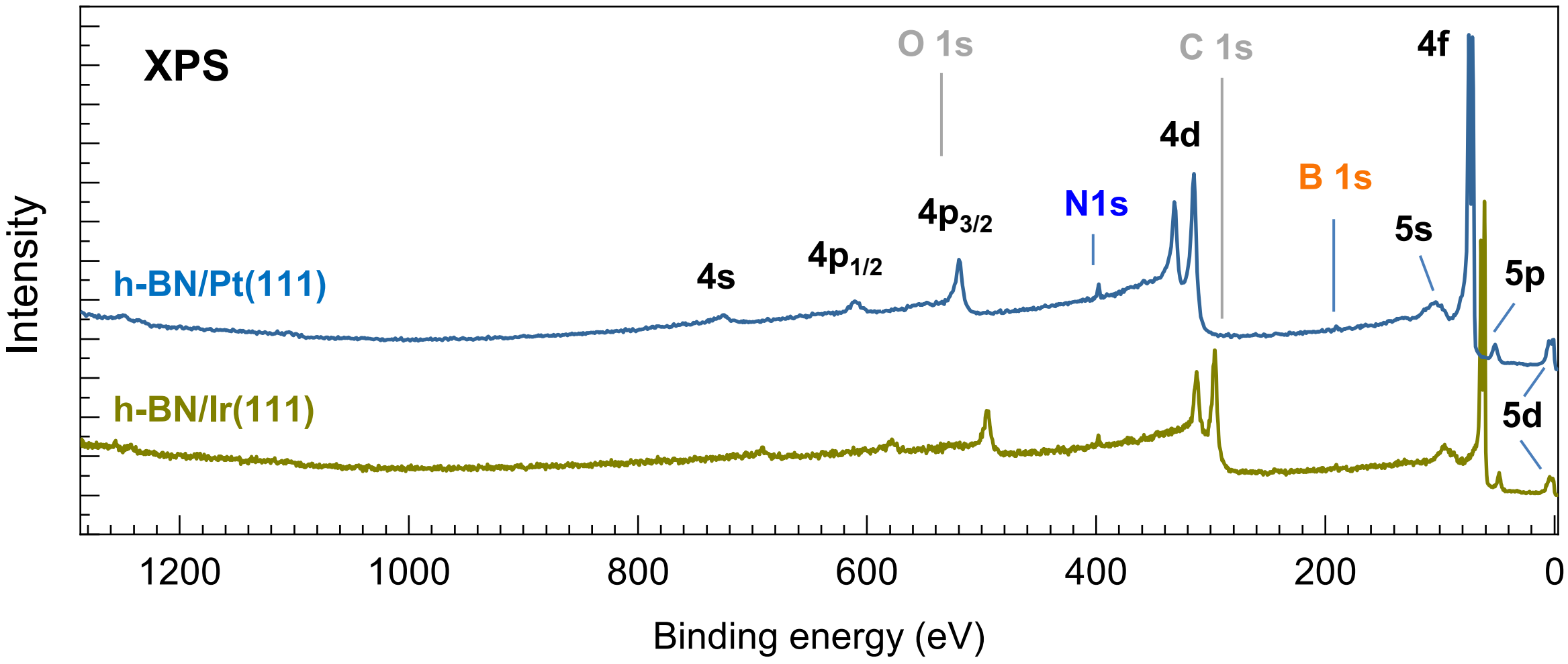

**Figure S2. Overview XPS spectra of monolayer h-BN on metal substrates.** Wide-range XPS spectra measured for monolayer h-BN grown on Pt(111) and Ir(111) at normal emission. The spectra exhibit the characteristic core-level peaks of the substrate metals together with the B 1s and N 1s signals originating from the h-BN overlayer. No additional peaks attributable to contaminant species such as oxygen (~531 eV) or carbon (~285 eV) are detected within the sensitivity of the measurement, indicating that high-quality h-BN layers were prepared on both substrates.

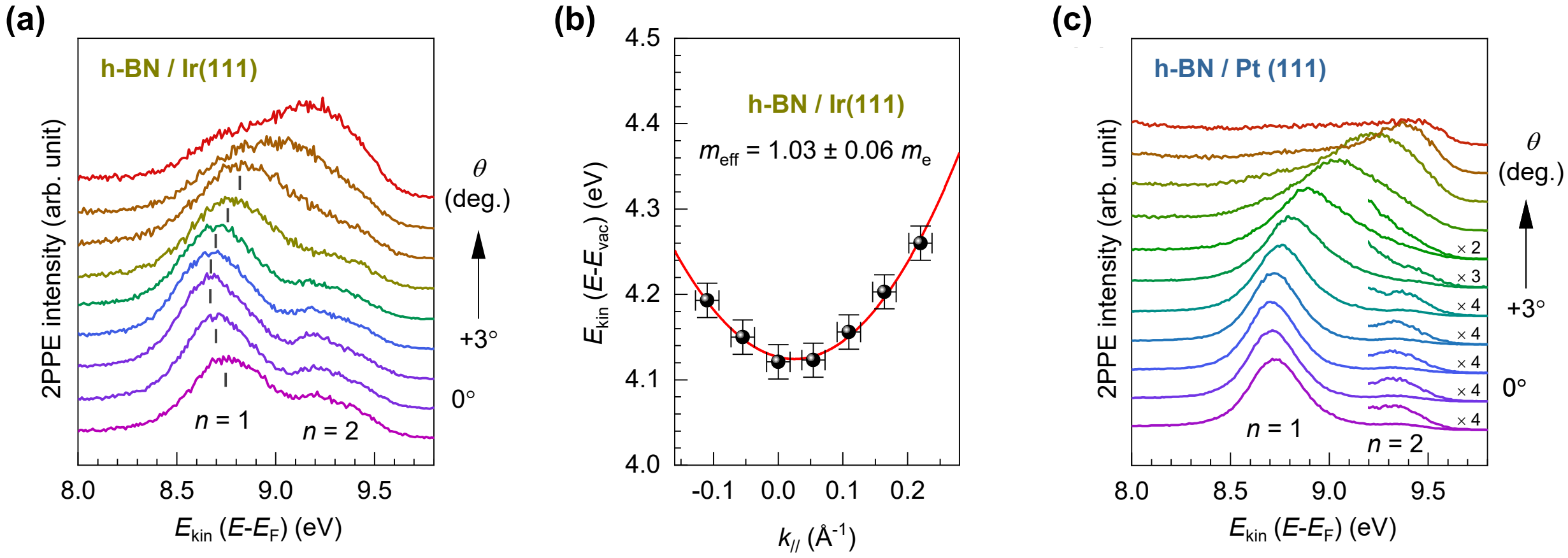

**Figure S3. Free-electron-like dispersion of IPS.** (a) Angle-resolved 2PPE spectra of monolayer h-BN/Ir(111) measured at $h\nu_{UV}$ = 4.77 eV. The spectra are recorded every 3° around the normal emission ($\theta$ = 0°) as denoted in the figure. (b) Extracted peak energy of the $n$ = 1 IPS on h-BN/Ir(111) plotted as a function of the parallel momentum, $k_{||}$. The $k_{||}$ was calculated using $k_{||} = (2m_e E_{\text{kin}})^{1/2} \sin(\theta) / \hbar$, where $E_{\text{kin}}$ is the photoelectron kinetic energy with respect to the vacuum level ($E_{\text{vac}}$), $m_e$ is the electron mass, and $\theta$ is the photoemission angle with respect to the surface normal. The data are fitted with a quadratic free-electron-like dispersion, yielding an effective mass of $m_{eff}$ = (1.03 ± 0.06) × $m_e$. (c) Corresponding 2PPE spectra for h-BN/Pt(111) recorded at $h\nu_{UV}$ = 4.77 eV. The $n$ = 1 and $n$ = 2 image potential states (IPS) exhibit systematic energy shifts with increasing emission angle, demonstrating free-electron-like dispersion. The nearly identical dispersion behavior on both substrates confirms that the static IPS electronic structure is largely preserved despite substrate-dependent structural corrugation.

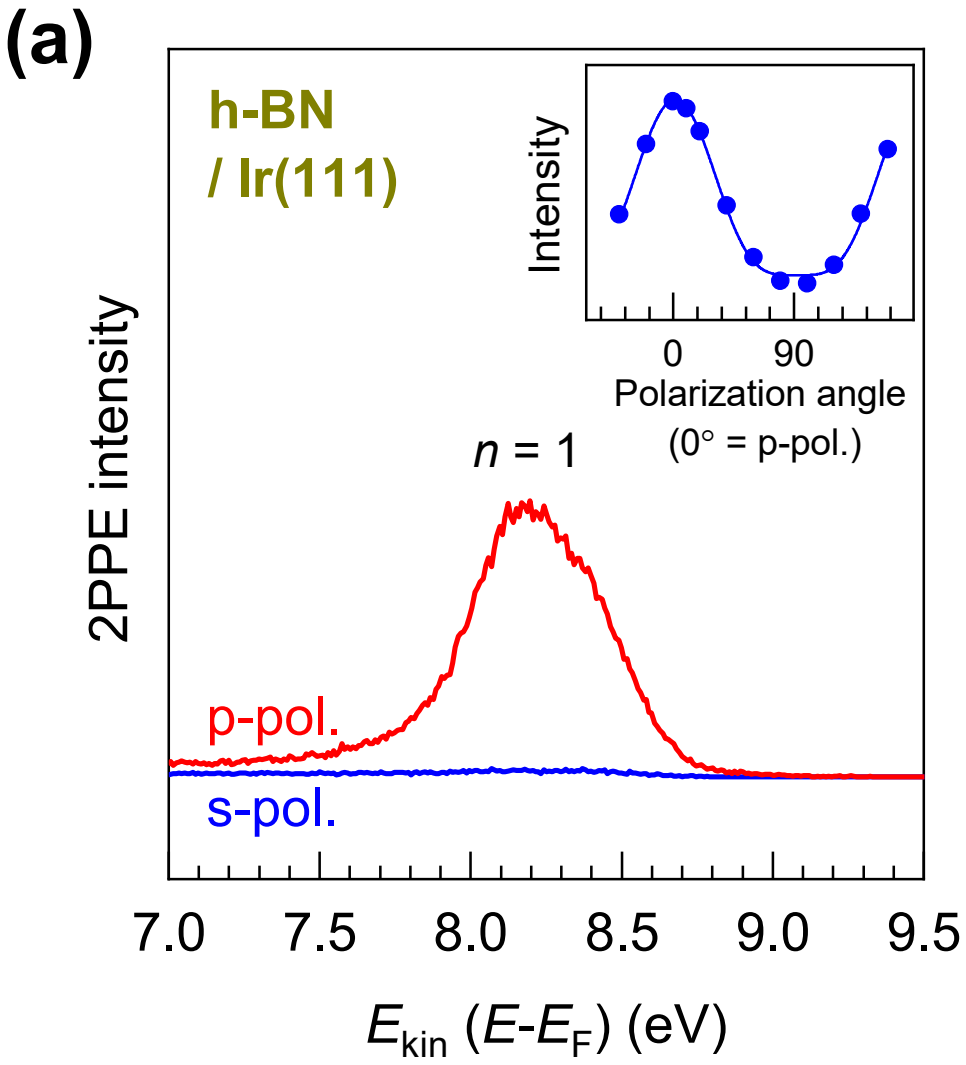

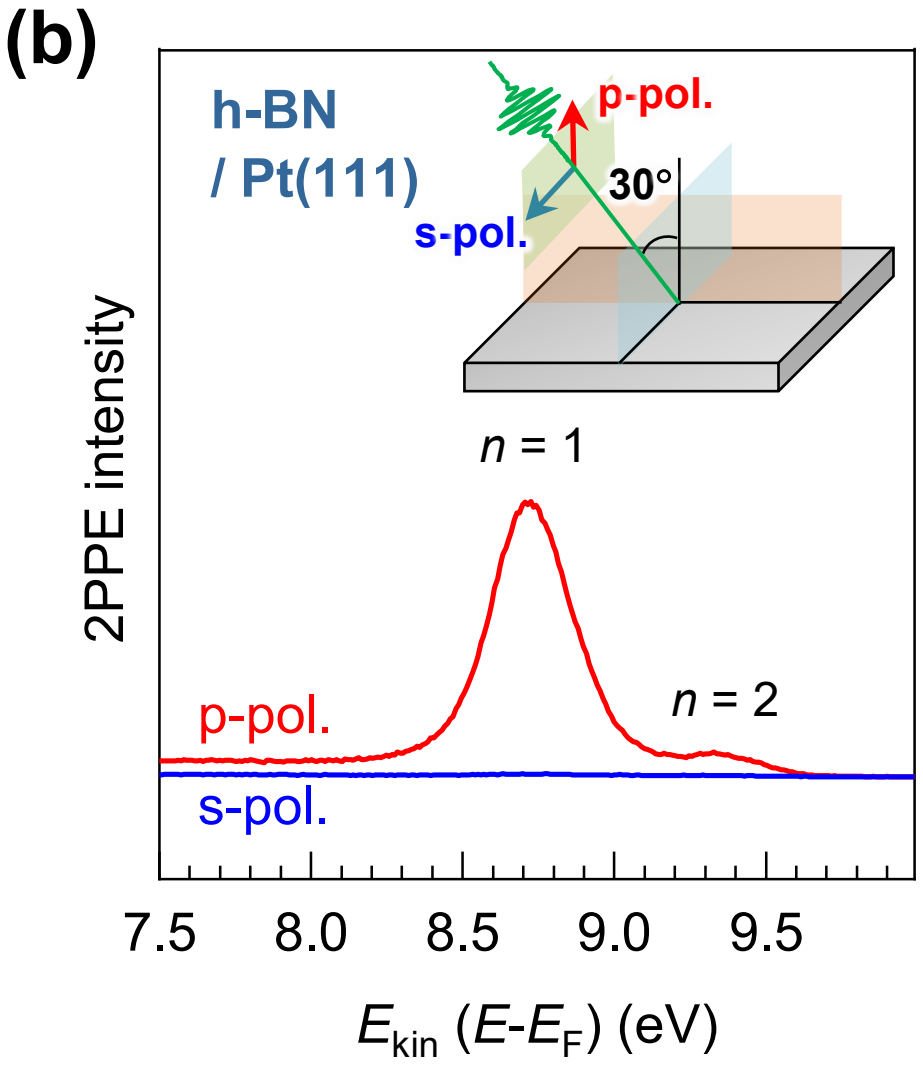

**Figure S4. Polarization dependence of the image potential states.** (a,b) Polarization-dependent 2PPE spectra of (a) h-BN/Ir(111) ($h\nu_{UV}$ = 4.33 eV) and (b) h-BN/Pt(111) ($h\nu_{UV}$ = 4.77 eV). In p-polarization, the electric field vector possesses a finite surface-normal component, whereas in s-polarization the electric field lies entirely parallel to the surface and therefore lacks a perpendicular component. Schematic illustration of the optical geometry is shown in the inset in (b). On both systems, the $n$ = 1 IPS intensity is strongly enhanced under p-polarized excitation (upper), while the signal is mostly suppressed for s-polarization (lower). The IPS intensity follows a $\cos^4(\varphi)$ dependence on the polarization angle φ (0° = p-polarization, see inset in (a)), consistent with a two-photon excitation process involving a surface-normal electric field component. The polarization selectivity for p-polarization is characteristic of the IPS, where the IPS electron is confined in the surface-normal direction while remaining free parallel to the surface, supporting the assignment of the observed features to IPS.